\documentclass[11pt]{article}
\usepackage{hyperref}
\pdfoutput=1
\begin{document}
\title{A Nanoaquarium for {\it in situ} Electron\\ Microscopy in Liquid Media}
\author{Joseph M. Grogan and Haim H. Bau \\
\\\vspace{6pt} Department of Mechanical Engineering and Applied Mechanics, \\ University of Pennsylvania, Philadelphia, PA 19104, USA}
\maketitle
\begin{abstract}
The understanding of many nanoscale processes occurring in liquids such as colloidal crystal formation, aggregation, nanowire growth, electrochemical deposition, and biological interactions would benefit greatly from real-time, {\it in situ} imaging with the nanoscale resolution of transmission electron microscopes (TEMs) and scanning transmission electron microscopes (STEMs). However, these imaging tools cannot readily be used to observe processes occurring in liquid media without addressing two experimental hurdles: sample thickness and sample evaporation in the high vacuum microscope chamber. To address these challenges, we have developed a nano-Hele-Shaw cell, dubbed the nanoaquarium \cite{GroganBauJMEMS,GroganBauUSNCTAM2010}. The device consists of a hermetically-sealed, 100 nm tall, liquid-filled chamber sandwiched between two freestanding, 50 nm thick, silicon nitride membranes. Embedded electrodes are integrated into the device for sensing and actuation. To demonstrate the capabilities of the device, this fluid dynamics video features particle motion and aggregation during {\it in situ} STEM imaging of nanoparticles suspended in liquids. The first solution contains 5 nm gold particles, 50 nm gold particles and 50 nm polystyrene particles in water. The second solution contains 5 nm gold particles in water. The imaging was carried out with a FEI Quanta 600 FEG Mark II with a STEM detector. The microscope was operated at 20-30 kV. In the footage of the multi-particle solution, note that the 50 nm gold particles prominently decorate the clusters and are clearly distinguished from the other particles. In the footage of the 5 nm gold particles, diffusion-limited aggregation is observed. Individual particles and small clusters are seen diffusing throughout the field of view, bumping into each other and bonding irreversibly to form a fractal structure. The rate of aggregation and the fractal dimension of the aggregates are consistent with light scattering measurements, indicating that the electron beam does not greatly alter the observed phenomenon \cite{GroganBauUnpub2010}.
\end{abstract}

\end{document}